\documentclass{preprint}

\usepackage{amsmath,amsthm,mathtools}
\usepackage{amssymb}
\usepackage{iftex}
\ifPDFTeX\usepackage[T1]{fontenc}\usepackage{lmodern}\fi
\usepackage{bm}
\usepackage{microtype}
\usepackage{booktabs}
\usepackage{needspace}
\usepackage[round]{natbib}

\newtheorem{proposition}{Proposition}

\definecolor{linkblue}{RGB}{20,60,140}
\hypersetup{allcolors=linkblue}

\makeatletter
\renewcommand\section{\@startsection{section}{1}{\z@}%
  {-3.5ex\@plus -1ex \@minus -.2ex}{1.2ex \@plus .2ex}%
  {\reset@font\large\bfseries\raggedright}}
\renewcommand\subsection{\@startsection{subsection}{2}{\z@}%
  {-3.25ex\@plus -1ex \@minus -.2ex}{0.8ex \@plus .2ex}%
  {\reset@font\normalsize\itshape\raggedright}}
\renewcommand\paragraph{\@startsection{paragraph}{4}{\z@}%
  {2.2ex \@plus .6ex \@minus .2ex}{-1em}%
  {\normalfont\normalsize\bfseries}}
\makeatother

\newcommand{\R}{\mathbb{R}}
\newcommand{\E}{\mathbb{E}}
\newcommand{\Ncal}{\mathcal{N}}
\newcommand{\thetab}{\bm{\theta}}
\newcommand{\xb}{\bm{x}}
\newcommand{\vb}{\bm{v}}
\newcommand{\thknown}{\thetab_{\mathrm{s}}}
\newcommand{\thunknown}{\thetab_{\mathrm{u}}}
\newcommand{\cmask}{\bm{c}}

\hypersetup{pdftitle={Amortized conditional inference under changing parameter information in small-angle X-ray scattering},pdfauthor={Maria B{\aa}nkestad et al.}}

\begin{document}

\title{Amortized conditional inference under changing\\ parameter information in small-angle X-ray scattering}

\author{Maria Bånkestad$^{1,*}$, Sandra Barman$^2$, Erik Kaunisto$^2$, Eskil Andreasson$^3$, Smita Chakraborty$^4$, Shun Yu$^4$, Jerk Rönnols$^1$ and Sepideh Pashami$^1$}

\affil{$^1$RISE Research Institutes of Sweden, Stockholm, Sweden}

\affil{$^2$RISE Research Institutes of Sweden, Göteborg, Sweden}

\affil{$^3$Tetra Pak, Lund, Sweden}

\affil{$^4$RISE Research Institutes of Sweden, Lund, Sweden}

\email{maria.bankestad@ri.se}

\begin{abstract}
Many scientific measurements do not uniquely determine the physical system that produced them. Different parameter configurations may explain the same observations, and additional measurements or physical assumptions can change which explanations remain plausible. We develop an amortized conditional inference method to examine these alternatives as the available information changes. Using masked flow matching, the method approximates joint conditional posteriors when different parameters are supplied exactly or with uncertainty. This allows repeated analysis of the same measurement without retraining, while reusing its encoded representation. We can also examine the consequences of assumed parameter distributions and separately assess whether supplied information is compatible with the observation-only posterior. We validate the learned conditionals against analytic posteriors and compare posterior accuracy and inference cost in a simulated scattering benchmark, where the compressed architecture performs better than joint attention on both measures. We apply the method to small-angle X-ray scattering (SAXS) of cellulose fibrils in paper, examining structural ambiguities in simulated measurements and the effects of physical assumptions on experimental measurements of a paper during hydration. At high humidity, the inferred effective fibril size distribution broadens. Assuming that this distribution remains unchanged instead shifts the posterior towards increased interfibrillar separation, although both interpretations reproduce the measured curves similarly. The compatibility analysis identifies conserved parameter values that are difficult to reconcile with the wettest measurements.
\end{abstract}


\section{Introduction}
\label{sec:intro}

Many scientific measurements do not uniquely determine the physical system that produced them. In small-angle X-ray scattering (SAXS) \citep{glatter1982saxs}, for example, the detector records the intensity of scattered X-rays but not their phase, limiting the structural information that can be recovered. The measured intensity depends on the material's spatial organization and the differences in scattering-length density between its components, known as scattering contrast. Different combinations of these properties can produce nearly indistinguishable curves. A good fit therefore does not establish which structure describes the material.

\begin{figure}[t]
  \centering
  \includegraphics[width=0.95\textwidth]{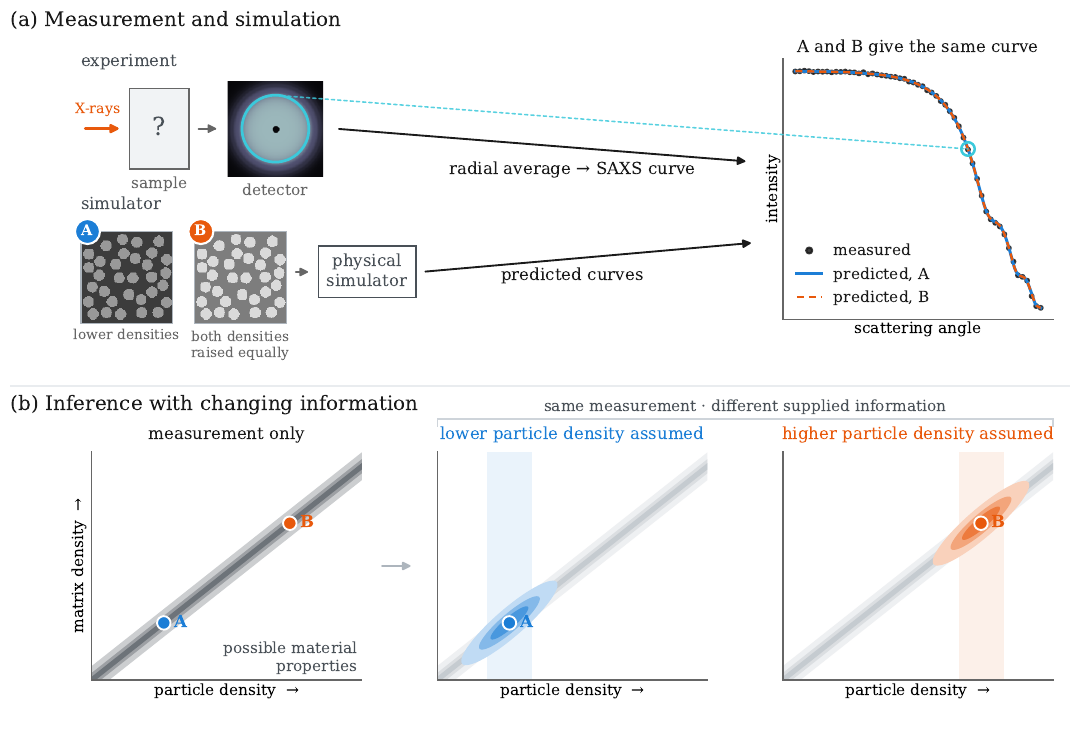}
  \caption{Conceptual illustration of simulator-based inference with changing parameter information.
\textbf{(a)}~In small-angle X-ray scattering (SAXS), an X-ray beam interacts with the sample, and the scattered intensity recorded on a detector is averaged around each ring into a scattering curve. A physical simulator predicts curves for proposed material configurations.
Configurations A and B have the same structure, but both phase densities of B are higher by the same amount (shared grey scale; lighter is denser). Because the scattering depends only on the density difference between the phases, A and B give the same curve, and the measurement cannot distinguish them.
\textbf{(b)}~Instead of selecting a single fitted solution, probabilistic inference represents the possible parameter values through a posterior distribution under the assumed model. Here it extends along a line of constant density difference, and A and B are two of its points. Supplying uncertain information about the particle density (shaded bands) changes which parts of this distribution are favored, although the measurement is unchanged. The material configurations and distributions are schematic.\looseness-1}
  \label{fig:concept}
\end{figure}

Interpreting such measurements requires understanding what they constrain under the assumed physical model. Some parameters may be well determined, others only through combinations with other parameters, and some may remain unresolved. A change in a fitted parameter does not necessarily mean that the measurement requires that change. For example, when studying cellulose hydration, we may want to know whether changes in the scattering curves require changes in fibril dimensions, or whether they can be explained while those dimensions remain unchanged. We may also want to examine how independent information about one property changes the inferred values of others, while retaining uncertainty in that information. These questions require examining how the joint parameter distribution changes with the supplied information, while also asking whether the measurement supports that information (Fig.~\ref{fig:concept}).

Simulation-based inference (SBI) addresses inverse problems where a forward simulator is available but the likelihood is difficult to evaluate \citep{beaumont2002abc,sisson2018handbook,cranmer2020frontier}. Rather than returning a single fitted solution, it estimates a posterior distribution over possible parameter values under the assumed model. Amortized methods learn reusable posterior approximations from simulated data, allowing new observations to be analyzed without solving the inverse problem from scratch \citep{papamakarios2016fast,greenberg2019apt}. Masked models further allow different subsets of parameters to be inferred within one trained network \citep{gloeckler2024allinone,nautiyal2026oneflowsbi}.

In scientific analysis, however, the questions often change even when the measurement remains the same. An analyst may first examine which structures are consistent with the measurement, then introduce information from another experiment or explore how the inferred structure changes under different physical assumptions. Each question requires a different joint conditional posterior, and the analysis may involve many such queries. This motivates a method that can handle changing parameter information, including uncertain values, without retraining or repeatedly encoding the observation.

We develop an amortized conditional inference method for this setting. Using masked flow matching, a single model approximates joint conditional posteriors when different parameters are supplied exactly or with uncertainty. The observation is encoded separately from the parameter information, allowing its representation to be reused across queries. We also distinguish between uncertain evidence, which updates the inferred parameter distribution, and assumed parameter distributions, whose consequences can be examined without updating them using the current measurement. Since a conditional posterior can be obtained even for poorly supported parameter information, we assess compatibility with the observation-only posterior separately. We validate the learned conditionals against analytic posteriors and compare posterior accuracy and inference cost in a simulated scattering benchmark.

We apply the method to simulated and experimental cellulose SAXS. Using simulated measurements with known parameters, we first examine which material properties can be inferred, which ambiguities remain, and how additional parameter information changes the posterior. We then analyze nine measurements of a paper, a network of cellulose fibres, taken during hydration, examining how the inferred structural changes depend on assumptions about conserved fibril properties. At high humidity, the observation-only inference suggests a broader effective fibril size distribution. Assuming that this distribution remains unchanged instead shifts the inference towards greater interfibrillar separation, although both interpretations reproduce the measured curves similarly. The analysis illustrates how changing physical assumptions can lead to different interpretations of the same measurements, while the compatibility analysis identifies conserved parameter values that are difficult to reconcile with the wettest measurements.

\section{Method}
\label{sec:method}
\begin{figure*}[t]
  \centering
  \includegraphics[width=\textwidth]{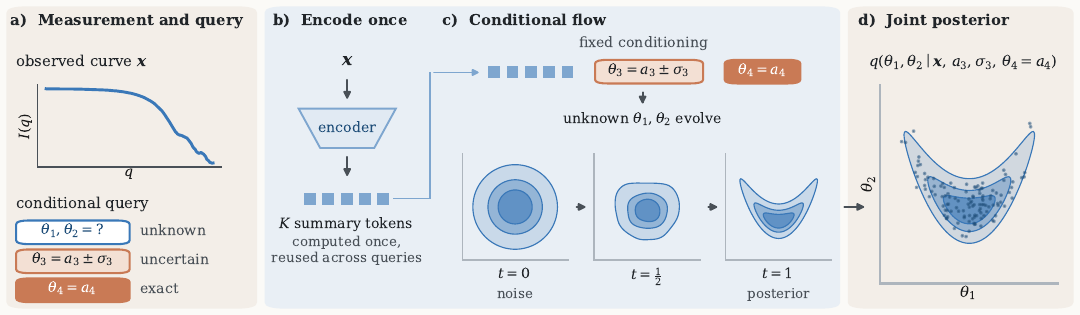}
  \caption{Overview of the conditional inference procedure.
  \textbf{(a)}~A measured curve $\xb$ is supplied together with a query. In the
  example, $\theta_1$ and $\theta_2$ are inferred jointly, $\theta_3$ is supplied
  with uncertainty $(a_3,\sigma_3)$, and $\theta_4$ is supplied exactly as
  $a_4$.
  \textbf{(b)}~The observation is compressed into $K$ summary tokens that are
  shared across queries, while the parameter tokens encode the current query.
  \textbf{(c)}~A conditional flow transforms noise into samples of the unknown
  parameters while the supplied information is held fixed as conditioning
  input.
  \textbf{(d)}~The result is the joint conditional posterior
  $q(\theta_1,\theta_2\mid\xb,a_3,\sigma_3,\theta_4=a_4)$.
  For uncertain information, the underlying parameter value is marginalized
  rather than fixed.\looseness-1}
  \label{fig:overview}
\end{figure*}

We consider repeated inference from the same measurement, with different parameters supplied exactly, supplied with uncertainty, or left unknown. Using simulated parameter–observation pairs, we train a single model to approximate the corresponding conditional posteriors. We encode the observation once and reuse its representation for different queries (Fig.~\ref{fig:overview}).

\subsection{Conditional inference setup}
\label{sec:setup}

We assume a simulator that maps physical parameters
$\thetab\in\R^d$ to observations $\xb\in\R^m$, together with a prior
$p(\thetab)$. The likelihood need not be available in closed form; we only
require that the simulator can generate joint samples $(\thetab,\xb)$.

For each query, every parameter has one of three roles:
\begin{itemize}
\item \emph{unknown}: inferred jointly with the other unknown parameters;
\item \emph{supplied exactly}: fixed to a specified value;
\item \emph{supplied with uncertainty}: represented by a reported value
$a_i$ and uncertainty $\sigma_i$.
\end{itemize}

Let $U$, $T$, and $E$ denote the sets of unknown, uncertain, and exactly
supplied parameters, respectively. The target distribution is
\begin{equation}
  p\!\left(
    \thetab_U
    \mid
    \xb,\,
    \bm{a}_T,\bm{\sigma}_T,\,
    \thetab_E=\bm{a}_E
  \right).
  \label{eq:mixed-query}
\end{equation}

The unknown parameters are inferred jointly. If no parameters are supplied, Eq.~\eqref{eq:mixed-query} reduces to the observation-only posterior.

\subsection{Masked conditional flow matching}
\label{sec:masking}
Different queries require inference over different subsets of parameters. We use masked conditional flow matching \citep{lipman2023flow} to learn the corresponding conditional distributions within a single model. Flow matching learns a velocity field that transforms Gaussian noise into posterior samples. Training uses parameter--observation pairs $(\thetab,\xb)$ generated by the simulator. For each pair, we draw Gaussian noise $\bm{\epsilon}\sim\Ncal(\bm{0},\bm{I})$ and define the linear path
\begin{equation}
\thetab_t
=
(1-t)\bm{\epsilon}
+
t\thetab,
\qquad
t\in[0,1].
\label{eq:flow-path}
\end{equation}
The target velocity along this path is $\thetab-\bm{\epsilon}$, which the model learns from $(\thetab_t,t,\xb)$. At inference, samples are generated by integrating the learned velocity field from $t=0$ to $t=1$, starting from Gaussian noise.

To condition on an exactly supplied subset of parameters, we introduce a
binary mask $\cmask\in\{0,1\}^d$. A supplied parameter has $c_i=1$ and is
kept fixed, whereas an unknown parameter has $c_i=0$ and is generated by the
flow. The state presented to the network is
\begin{equation}
    \widetilde{\thetab}
    =
    \cmask\odot\thetab
    +
    (\bm{1}-\cmask)\odot\thetab_t .
    \label{eq:masked-state}
\end{equation}
The mask changes which parameters the flow generates: supplied values remain fixed, while the unknown parameters evolve jointly, conditional on both the measurement and the supplied values. The flow-matching loss is evaluated only on the unknown
coordinates; the full objective is given in
Appendix~\ref{app:masked-objective}.

We also provide the mask to the network. During training, we sample masks
corresponding to the conditioning patterns of interest. The unknown parameters
are modeled jointly rather than independently. As a result, the velocity of one
parameter may depend on the observation, the supplied parameter values, and the
current values of the other unknown parameters. The model can therefore
capture dependencies among the inferred parameters.

At the population optimum, masked training recovers the conditional velocity field for each conditioning pattern represented during training. This does not guarantee equal finite-sample accuracy across patterns. The result assumes base noise sampled independently of the parameters, as used throughout this work; we give the formal statement and proof in Appendices~\ref{app:masked-objective} and~\ref{app:proofs}.

\subsection{Conditioning on uncertain statements}
\label{sec:soft}

When a parameter is supplied with uncertainty, we treat the reported value $a_i$ as a noisy observation of the underlying parameter $\theta_i$, with likelihood
\[
p(a_i\mid\theta_i,\sigma_i).
\]
We use a Gaussian likelihood, where $\sigma_i$ describes the uncertainty of the reported value rather than prior uncertainty about $\theta_i$. The reported value provides information about the parameter without fixing its underlying value. It may come from an independent measurement or other uncertain information about the physical state.

During training, uncertain statements are generated as
\begin{equation}
    a_i = \theta_i + \sigma_i\eta_i,
    \qquad
    \eta_i\sim\Ncal(0,1),
    \label{eq:statement-corruption}
\end{equation}
with $\sigma_i$ sampled from a prescribed range in the normalized parameter
coordinates used for training (Appendix~\ref{app:impl}). Exact values are handled by
the exact-conditioning mechanism described in Section~\ref{sec:masking}, so
exact and uncertain information can be queried with the same trained model.
The target in Eq.~\eqref{eq:mixed-query} then becomes
\begin{align}
    &p\!\left(
        \thetab_U
        \mid
        \xb,\bm{a}_T,\bm{\sigma}_T,
        \thetab_E=\bm{a}_E
    \right)
    =
    \int
    p\!\left(
        \thetab_U
        \mid
        \xb,\thetab_T,
        \thetab_E=\bm{a}_E
    \right)
    p\!\left(
        \thetab_T
        \mid
        \xb,\bm{a}_T,\bm{\sigma}_T,
        \thetab_E=\bm{a}_E
    \right)
    d\thetab_T .
    \label{eq:statement-marginal}
\end{align}
The uncertain parameters \(\thetab_T\) are not fixed. Their distribution is updated using both the measurement \(\xb\) and the supplied statements \((\bm{a}_T,\bm{\sigma}_T)\), and then marginalized to obtain the joint posterior over the remaining unknown parameters \(\thetab_U\).

At the population optimum of this training objective, with statements
generated by Eq.~\eqref{eq:statement-corruption}, the model recovers this
marginalized conditional distribution. The formal result and proof are given in
Appendices~\ref{app:uncertain-objective} and~\ref{app:proofs}.

Two assumptions are important when the supplied information comes from an
external measurement. First, the external measurement is assumed conditionally
independent of $\xb$ given the underlying parameters. Shared systematic errors
between the two measurements would violate this assumption. Second, we use a
factorized statement likelihood across coordinates. Correlated external
measurements would instead require a multivariate statement model.

\paragraph{Conditioning on an assumed distribution.}
Sometimes we want to examine the consequences of a physical assumption rather than update the inference with an additional measurement. For example, we may assume that certain properties remain unchanged between measurements, although their values are uncertain. We can then use their joint posterior from the first measurement as an assumed distribution \(\pi(\theta_E)\), retaining uncertainty and correlations between the conserved parameters. We draw values \(s\sim\pi(\theta_E)\), condition exactly on each draw, and pool the resulting samples:
\begin{equation}
\int
p\!\left(\thetab_U \mid \xb, \thetab_E=\bm{s}\right)
\pi(\bm{s})\, d\bm{s}.
\label{eq:assumed-mixture}
\end{equation}
Unlike the uncertain statements in Eq.~\eqref{eq:statement-marginal}, the assumed distribution \(\pi\) is not updated by the current measurement. The same sampled values are used when conditioning on each measurement in the series, allowing the remaining parameters to be compared under a common assumption.

\paragraph{Range of supplied uncertainty.}
Conditional inference may be unreliable for reported values or uncertainty levels poorly represented during training. We therefore restrict the supplied uncertainty scales to the training range and assess compatibility separately from conditional inference.

\subsection{Compatibility of supplied statements}
\label{sec:compat}
Conditioning on a supplied value does not establish that the value is supported by the measurement. We therefore assess its compatibility separately, using the observation-only posterior $q(\thetab\mid\xb)$.

For a supplied statement, we average its likelihood over samples from the observation-only posterior. The resulting score is compared with scores of reference statements generated from the same posterior using the corresponding statement uncertainty. We report the percentile of the supplied score within this reference distribution. Low percentiles indicate that the statement is unusual under the observation-only posterior, and we flag values below $0.01$. Because the score uses the statement likelihood, the percentile also depends on the uncertainty assigned to the statement: a broader statement with the same value can be compatible where a narrow one is flagged.

The compatibility percentile is a diagnostic, not a formal significance level or the probability that a statement is correct. The finite-sample calculation is described in
Appendix~\ref{app:evaluation}.

\subsection{Architecture}
\label{sec:arch}
To avoid processing the same observation for every conditional query, we separate its representation from the changing parameter information. An encoder compresses the observation into shared summary vectors, while parameter tokens represent the current query and attend to these summaries when predicting the flow.

\paragraph{Parameter tokens.}
Each parameter has a token encoding its identity and role in the current query. For unknown parameters, the token contains the current flow state $\theta_{i,t}$; for exactly supplied parameters, the supplied value; and for uncertain parameters, the reported value and uncertainty $(a_i,\sigma_i)$. The latent value of an uncertain parameter is not supplied to the network, consistent with Eq.~\eqref{eq:statement-marginal}.

\paragraph{Observation summary.}
The encoder maps the observation to $K$ summary vectors,
\[
    \bm{s}_{1:K} = \mathrm{Enc}(\xb),
    \qquad K \ll m.
\]

The parameter tokens access these summaries through cross-attention and interact with each other through self-attention \citep{vaswani2017attention}. The network also receives the flow time $t$, and a separate output head predicts each parameter's velocity. Appendix~\ref{app:impl} provides architectural details.

\subsection{Inference}
\label{sec:sampling}

At inference, we initialize the unknown parameters from Gaussian noise and integrate the learned velocity field from $t=0$ to $t=1$ using an ODE solver. The supplied information remains fixed during integration, while the unknown parameters evolve jointly.

We handle parameter bounds with a smooth invertible transformation to an
unconstrained space rather than by clipping (Appendix~\ref{app:impl}).
Because the observation is unchanged across repeated queries, its encoded
representation can be reused across solver steps and queries. The corresponding
complexity analysis and timing protocol are given in
Appendix~\ref{app:impl}.

\subsection{Evaluation without a reference posterior}
\label{sec:eval}

For measured data, the true posterior is unknown, so we cannot directly assess whether the inferred distribution is correct. Simulated test cases provide an alternative: the generating parameters are known even when the corresponding posterior cannot be calculated analytically. We use these samples to evaluate exact conditional inference without requiring a reference posterior in closed form.

Consider a held-out pair $(\thetab^*,\xb)$ drawn from the simulator. We choose a
conditioning mask, supply the corresponding coordinates
$\thknown^*$, and query
\[
q(\thunknown \mid \xb,\thknown^*).
\]
Since the held-out parameters and observation were generated jointly, the
remaining parameter values $\thunknown^*$ are a sample from the true
conditional distribution,
\[
p(\thunknown \mid \xb,\thknown^*).
\]
This allows us to evaluate whether the learned posterior assigns appropriate
probability to draws from the true conditional distribution, without requiring
that distribution in closed form. We use exactly supplied parameter values here to evaluate conditional posterior accuracy independently of the treatment of uncertain statements.

We first assess calibration. Marginal calibration is evaluated with conditional
simulation-based calibration \citep{talts2018sbc}, and joint expected coverage
with TARP \citep{lemos2023tarp}. Coverage tests whether posterior uncertainty has
the expected frequency, but it does not fully characterize the shape or
dependence structure of the conditional distribution.
We therefore also evaluate the conditional log density assigned to the held-out
unknown parameters,
\begin{equation}
\mathcal{S}(q)
=
\E_{(\thetab^*,\xb),\cmask}
\left[
\log q(\thunknown^* \mid \xb,\thknown^*)
\right].
\label{eq:logprob-score}
\end{equation}
The flow provides this density through change of variables. All models are
evaluated on the same held-out observations and conditioning masks. The
conditional log score is strictly proper: in expectation it is maximized by the
true conditional distribution, and differences in score correspond to
differences in expected KL divergence. The formal result is given in
Appendix~\ref{app:conditional-score}.

Both coverage and conditional log score evaluate the learned posterior under the simulator and prior. Neither tests whether the simulator adequately describes the physical experiment.

\section{Experiments}
\label{sec:experiments}

We use two test problems: a Gaussian model with analytically known conditionals and the cellulose SAXS simulator.

\subsection{Tasks and evaluation setup}
\label{sec:exp-setup}

We use two complementary test problems. The Gaussian task
has analytically known conditional posteriors, allowing us
to evaluate exact and uncertain conditioning against a
known target. The cellulose SAXS simulator provides a
nonlinear inverse problem with parameter ambiguities and
no analytically available posterior. We use it to compare
conditional posterior accuracy and computational cost
across architectures.

\paragraph{Gaussian task.}
The Gaussian task has five correlated parameters observed through a lower-dimensional linear projection.

\begin{figure}[t]
  \centering
  \includegraphics[width=0.9\textwidth]{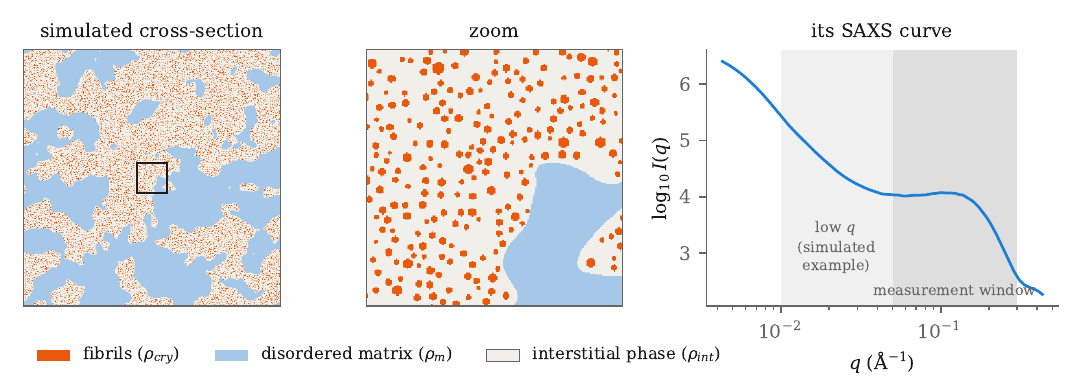}
  \caption{The scattering simulator, in the configuration used in the
  application. \emph{Left:} a simulated
  cross-section with fibrils, disordered matrix and interstitial phase;
  \emph{middle:} zoom of the marked region; \emph{right:} the simulated SAXS
  curve. Shaded: the $q$-range used for the measured curves
  ($0.05$--$0.30$~\AA$^{-1}$, dark) and its low-$q$ extension used for the
  simulated example (light). The parameters are those of the simulated example
  analyzed in Fig.~\ref{fig:app2-posterior}: fibril radius $r_f=1.29$~nm, size
  spread $c_f=0.20$, nominal fibril count $N_f=19{,}950$, minimum separation
  $s_{\min}=0.41$~nm, matrix volume fraction $\phi=0.41$, matrix correlation
  length $\ell_m=2.37$ (simulation units), and phase densities
  $\rho_{cry},\rho_m,\rho_{int}=0.80,0.72,0.64$. The cross-section is one
  realization with these parameters, and the curve is the average over six
  realizations.}
  \label{fig:simulator}
\end{figure}

\paragraph{Scattering task.}
The scattering task uses a physical simulator of two-dimensional
cross-sections perpendicular to parallel cellulose fibrils
(Fig.~\ref{fig:simulator}). We use the same physical model in the application
of Section~\ref{sec:app2}, with a more detailed representation of fibril
sizes. Each cross-section contains three phases: hexagonal cellulose fibrils, a disordered matrix surrounding them, and the remaining interstitial space. Each phase has a uniform scattering-length density, with sharp boundaries between phases. The spatial structure of the disordered matrix is generated from a thresholded Gaussian random field. The geometric
parameters are the fibril radius $r_f$, nominal fibril count $N_f$, minimum
fibril separation $s_{\min}$, disordered-matrix volume fraction $\phi$ (the
fraction of the cross-section occupied by the disordered matrix), and matrix
correlation length $\ell_m$. Three further parameters
specify the scattering-length densities of the fibrils ($\rho_{cry}$), the
matrix ($\rho_m$), and the interstitial phase ($\rho_{int}$), hereafter phase
densities.\looseness-1

The simulator represents some of the structural features included in analytical cellulose scattering models such as WoodSAS \citep{penttila2019woodsas,penttila2020bundling,penttila2021wateraccess}, but generates the fibrils and surrounding matrix explicitly in real space, with fibril centers not restricted to a lattice.

The benchmark uses eight parameters, with all fibrils having the same size.
The application uses a nine-parameter configuration that also describes the
spread of fibril sizes (Appendix~\ref{app:simulator}).

The simulated SAXS curve is obtained by azimuthally averaging the squared
Fourier transform of the two-dimensional density map
\citep{schmidt2007simulation,roding2022,bankestad2026saxs}, and averaging over six
independently generated structures. Because the three phases partition the
cross-section, the scattering at nonzero $q$ depends on density contrasts
rather than on absolute densities:
\begin{equation}
  I(q) = \bm{\delta}^\top S(q; g)\,\bm{\delta},\qquad
  \bm{\delta} = \begin{pmatrix}\rho_m-\rho_{cry}\\ \rho_{int}-\rho_{cry}\end{pmatrix}.
  \label{eq:contrast-form}
\end{equation}
Here $g$ denotes the geometric parameters and $S(q;g)$ their contributions to
the scattering (Appendix~\ref{app:simulator}). The
simulated $q$-axis is mapped to \AA$^{-1}$ by a fixed calibration, and
Table~\ref{tab:params} lists the parameters with their units and prior
ranges.

For the models in this section, the observation has dimension $m=67$ and the parameter vector dimension $d=8$. With $K=8$ observation-summary tokens, each solver step therefore operates on $16$ tokens rather than the $75$ tokens required by a joint parameter--observation representation.

\paragraph{Shared evaluation.}
Within each comparison, all models use the same held-out examples, conditioning patterns, and random draws. Scattering results are reported over three training seeds. The conditional log score measures how well the conditional density is estimated. Marginal $90\%$ coverage and the TARP joint-coverage error measure calibration. Recovery is the normalized RMSE between the posterior mean and the generating parameters. Appendix~\ref{app:impl} provides further details.

\subsection{Validation against known conditionals}
\label{sec:exp-synthetic}

We compare the learned Gaussian conditionals with the analytic posteriors for five levels of uncertainty in the supplied parameters (Table~\ref{tab:gauss-stmt}). We use a classifier two-sample test \citep[C2ST;][]{lueckmann2021benchmarking}, where an accuracy of $0.5$ indicates that the classifier cannot distinguish the two sets of samples better than chance.

\begin{table}[ht]
\centering
\caption{Validation of exact and uncertain conditioning on the Gaussian task.
The same trained model is compared with the analytic conditional posterior at five uncertainties.
C2ST accuracy of $0.5$ is chance level and lower KL is better. The compatibility row reports mean percentiles for statements generated from the analytic model; values below $0.01$ are flagged as incompatible.
Results are averaged over $12$ conditioning patterns.}
\label{tab:gauss-stmt}
\small
\setlength{\tabcolsep}{4pt}
\begin{tabular}{lccccc}
\toprule
uncertainty $\sigma$ & $0$ & $0.02$ & $0.05$ & $0.10$ & $0.15$ \\
\midrule
C2ST vs analytic
& $0.543 \pm 0.072$
& $0.513 \pm 0.021$
& $0.521 \pm 0.027$
& $0.542 \pm 0.035$
& $0.570 \pm 0.059$ \\
compatibility percentile
& $0.60$
& $0.55$
& $0.55$
& $0.56$
& $0.55$ \\
$\mathrm{KL}(p_{\mathrm{analytic}}\|q)$ (nat)
& $0.78 \pm 0.60$
& $0.22 \pm 0.26$
& $0.29 \pm 0.25$
& $0.45 \pm 0.30$
& $0.60 \pm 0.33$ \\
\bottomrule
\end{tabular}
\end{table}

C2ST accuracy ranges from $0.513$ to $0.570$
across the five uncertainty levels
(Table~\ref{tab:gauss-stmt}). These values indicate
that the learned and analytic distributions are
difficult to distinguish with the classifier at
the available sample size, but do not establish
that they are identical. The KL divergences range
from $0.22$ to $0.78$ nat, showing that
approximation errors remain.

We also test the compatibility diagnostic using $50$ deliberately conflicting statements with uncertainty $0.03$. All $50$ have compatibility percentiles below $10^{-3}$ and are flagged at the prespecified threshold of $0.01$. None of the compatible statements in Table~\ref{tab:gauss-stmt} is flagged.

The comparison of two statement-embedding designs gives similar C2ST results but a difference of $0.42\pm0.08$ nat in conditional log score (Table~\ref{tab:gauss-metrics}; Appendix~\ref{app:impl}). This suggests that the models differ in density estimation accuracy even when the classifier test does not distinguish them clearly. We therefore use conditional log scores alongside coverage diagnostics in the scattering benchmark.

\subsection{Observation compression and posterior accuracy}
\label{sec:exp-arch}

The compressed observation representation is useful only
if it retains the information needed for conditional
inference. On the scattering task, we therefore compare
models with $K\in\{4,8,16\}$ summary tokens against a
model that jointly attends over observation and parameter
tokens. We also include a no-attention baseline to examine
the role of interactions between parameter tokens.
All other architecture and training settings are matched.

\begin{table}[ht]
\centering
\small
\caption{Posterior accuracy on the scattering task (mean $\pm$ s.d.\ over three seeds; metrics as in Section~\ref{sec:exp-setup}).
Higher conditional log score is better; nominal marginal coverage is $0.90$; lower TARP and recovery error are better.}
\label{tab:saxs}
\begin{tabular}{lcccc}
\toprule
model & score & cov$_{90}$ & TARP err.\ & recovery \\
\midrule
compressed, $K=4$  & $7.52 \pm 0.13$ & $0.905 \pm 0.013$ & $0.012 \pm 0.006$ & $0.108 \pm 0.001$ \\
compressed, $K=8$  & $\mathbf{7.65 \pm 0.03}$ & $0.901 \pm 0.005$ & $0.013 \pm 0.003$ & $\mathbf{0.107 \pm 0.000}$ \\
compressed, $K=16$ & $7.58 \pm 0.04$ & $0.904 \pm 0.004$ & $0.013 \pm 0.003$ & $0.108 \pm 0.000$ \\
joint attention    & $6.95 \pm 0.03$ & $0.895 \pm 0.015$ & $0.009 \pm 0.005$ & $0.110 \pm 0.001$ \\
no attention       & $1.48 \pm 0.02$ & $0.864 \pm 0.004$ & $0.007 \pm 0.001$ & $0.152 \pm 0.000$ \\
\bottomrule
\end{tabular}
\end{table}
All three compressed models achieve higher conditional log scores than the joint-attention baseline (Table~\ref{tab:saxs}). The highest score is obtained with $K=8$ ($7.65\pm0.03$, compared with $6.95\pm0.03$ for joint attention). Marginal coverage and posterior-mean recovery are similar, and performance changes little between $K=4$ and $K=16$.

The no-attention baseline has the lowest TARP error, but its conditional log score is more than six nats below the best model, and its posterior-mean recovery is substantially worse. Good coverage alone is therefore insufficient to establish posterior accuracy.

\subsection{Computational cost of conditional inference}
\label{sec:exp-cost}

On an A100 GPU, drawing $1{,}000$ posterior samples takes $408$\,ms with the compressed model and $1{,}393$\,ms with joint attention, a $3.4\times$ speedup. The corresponding speedup on CPU is $7.1\times$.

\begin{figure*}[t]
  \centering
  \includegraphics[width=0.7\textwidth]{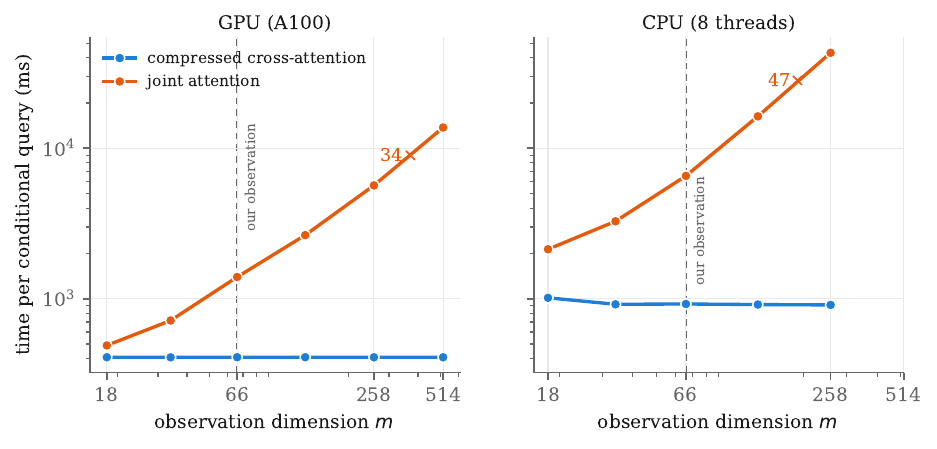}
  \caption{Time to answer one conditional query ($1{,}000$ samples) as the observation dimension grows. Ratios are marked at the largest $m$ measured;
  the dashed line marks $m=66$, the benchmark dimension closest to the
  observation of the trained models ($m=67$).
  Timings use untrained models, since forward cost does not depend on the weights.}
  \label{fig:cost}
\end{figure*}

To examine how this difference scales with observation size, we vary the observation dimension from $m=18$ to $514$ while keeping the rest of the architecture fixed (Fig.~\ref{fig:cost}). Query time changes little for the compressed model, whereas joint attention becomes increasingly expensive. At the largest tested dimension, joint attention is $33.7\times$ slower on the GPU.

These timings compare individual queries. Reusing the
encoded observation across repeated queries can provide
further savings, but this additional benefit is not
measured here.

\subsection{Exact conditioning after uncertain-statement training}
\label{sec:exp-hard}

We find no measurable loss in exact-conditioning accuracy when training with uncertain statements. At zero uncertainty, the conditional log score is $7.65\pm0.03$, compared with $7.59\pm0.12$ for training with exact statements alone. Coverage, recovery and query time are also similar across the three training seeds.

For the cellulose analysis in Section~\ref{sec:app2}, we use models trained on the nine-parameter simulator, which includes the spread of fibril sizes (Appendix~\ref{app:simulator}).

\section{Cellulose SAXS: what do the measurements tell us?}
\label{sec:app2}

We first analyze a simulated SAXS measurement with known parameters, followed by measurements of cellulose at different humidities. The simulated case allows us to examine structural ambiguities and the effects of additional parameter information before interpreting the measured series.

\subsection{A simulated measurement with known parameters}
\label{sec:app2-sim}

We simulate one SAXS curve from the material in Fig.~\ref{fig:simulator}, whose
generating parameters are listed in the caption, and add noise at the level of
the measured curves. We analyze the noisy curve over the full simulated range,
$0.01$--$0.30$~\AA$^{-1}$, and over the window used for the measured curves in
Section~\ref{sec:app2-series}, $0.05$--$0.30$~\AA$^{-1}$. Posterior samples
are pooled from three independently trained models
(Appendix~\ref{app:instrument}).

\begin{figure}[t]
  \centering
  \includegraphics[width=0.9\textwidth]{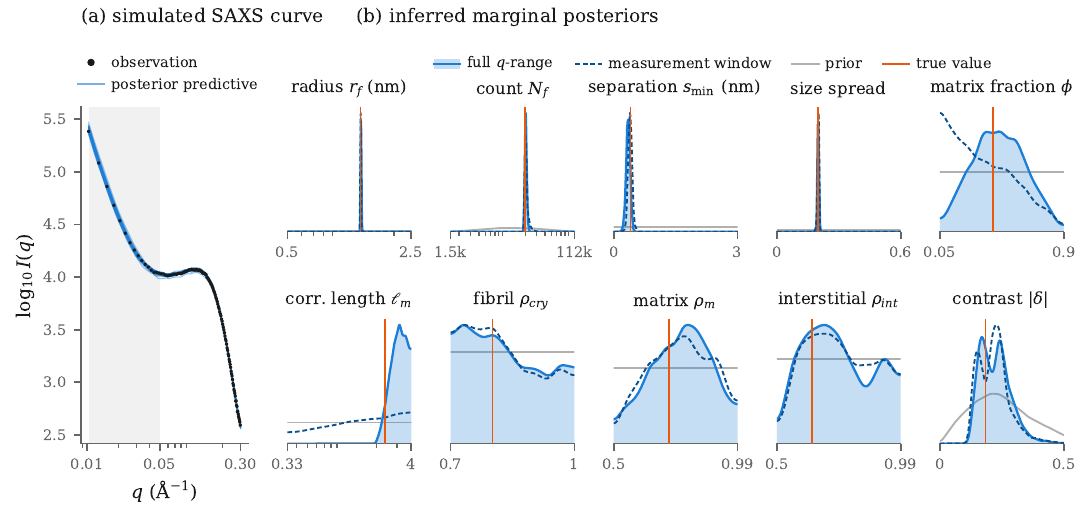}
  \caption{Posterior inference from a simulated SAXS measurement with known
  parameters.
  \textbf{(a)}~Noisy observation and 30 posterior-predictive curves. The shaded
  region marks low-$q$ data outside the window used for the measured curves.
  \textbf{(b)}~Marginal posteriors from the full simulated $q$-range (solid)
  and from the same observation restricted to the measurement window
  $0.05$--$0.30$~\AA$^{-1}$ (dashed). Grey: prior; orange: generating value.
  Each axis spans the prior range. Samples are pooled from three independently trained
  models.}
  \label{fig:app2-posterior}
\end{figure}

The simulated curve constrains the fibril radius, size spread, nominal count and minimum separation to narrow ranges (Fig.~\ref{fig:app2-posterior}b). The $90\%$ interval for the radius is $1.29$--$1.31$~nm. Excluding the low-$q$ region has little effect on these parameters, but broadens the interval for the matrix correlation length from $2.3$--$3.9$ to $0.4$--$3.7$ simulation units, nearly its full prior range of $0.33$--$4$. Appendix~\ref{app:instrument} gives the calibration results.

\begin{figure}[t]
  \centering
  \includegraphics[width=\textwidth]{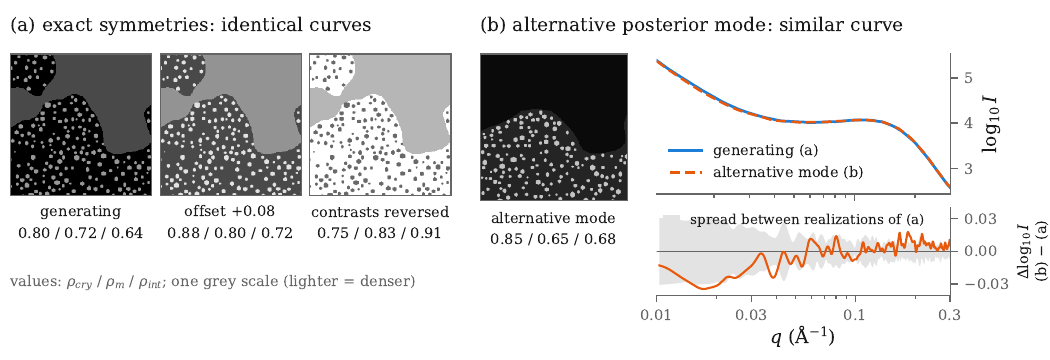}
  \caption{Different density configurations can produce identical or similar
  SAXS intensities. All cross-sections use the same grey scale; the values
  below them are the fibril, matrix and interstitial densities.
  \textbf{(a)}~The generating structure, the same structure with a common
  offset added to all three densities, and with all density contrasts reversed
  produce identical scattering intensities.
  \textbf{(b)}~An alternative configuration sampled from the inferred
  posterior, with different phase densities and geometry. Its predicted
  scattering is similar but not identical to that of the generating structure;
  the lower plot shows their difference against the variation between
  independent structural realizations of the generating material.\looseness-1}
  \label{fig:app2-invariance}
\end{figure}

The matrix volume fraction and phase densities remain poorly constrained. The density parameters also have exact symmetries: scattering depends on the three phase densities only through two contrasts  $\bm{\delta}$ (Eq.~\eqref{eq:contrast-form}). Adding a common offset to all three densities or reversing both contrasts leaves the scattering intensity unchanged (Fig.~\ref{fig:app2-invariance}a). These ambiguities cannot be resolved by reducing the measurement noise, because the predicted intensities are exactly the same.

The posterior also contains configurations with a different ordering of the matrix and interstitial densities and different fibril geometry (Fig.~\ref{fig:app2-invariance}b). Unlike the exact symmetries, these configurations produce slightly different scattering curves, mainly below $0.03$~\AA$^{-1}$ and above $0.15$~\AA$^{-1}$. The exact density symmetries represent structural non-identifiability, whereas the alternative configurations are difficult to distinguish under the simulated measurement conditions \citep{raue2009structural,transtrum2015sloppiness}.

\begin{figure}[t]
  \centering
  \includegraphics[width=0.95\textwidth]{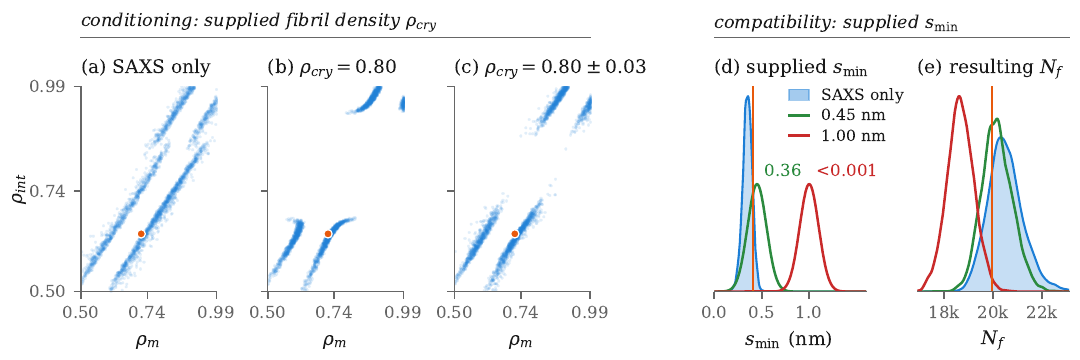}
  \caption{Conditional inference with exact and uncertain parameter
  information, for the simulated observation of
  Fig.~\ref{fig:app2-posterior} (full $q$-range).
  \textbf{(a--c)}~Joint posterior of the matrix and interstitial densities
  without additional information, with the generating fibril density
  supplied exactly, and with the same density supplied with standard deviation
  $0.03$.
  \textbf{(d)}~Measurement-only posterior of the minimum fibril separation
  (blue) and two supplied Gaussian statements with standard deviation
  $0.10$~nm, centred at $0.45$~nm (green) and $1.00$~nm (red; scaled for
  display), with their compatibility percentiles.
  \textbf{(e)}~The corresponding posteriors of the nominal fibril count, in the
  same colors. Orange: generating values.}
  \label{fig:app2-conditioning}
\end{figure}

We next examine how additional information about the fibril density affects these ambiguities. Without additional information, the joint posterior of the matrix and interstitial densities forms several elongated groups (Fig.~\ref{fig:app2-conditioning}a--c). Fixing the fibril density at its generating value, $\rho_{cry}=0.80$, removes the density-offset ambiguity. One group concentrates near the generating densities, but the others remain because of the contrast-sign and phase-ordering ambiguities. Supplying the fibril density with standard deviation $0.03$ instead broadens the groups along the offset direction.

We next consider two Gaussian statements about minimum fibril separation, each with standard deviation $0.10$~nm: one centered at $0.45$~nm, close to the generating value of $0.41$~nm, and another at $1.00$~nm, outside the observation-only posterior (Fig.~\ref{fig:app2-conditioning}d). Their compatibility percentiles are $0.36$ ($0.29$--$0.43$ across the three models) and below $0.001$, respectively; only the second is flagged as conflicting.\looseness-1

Both statements produce conditional posteriors (Fig.~\ref{fig:app2-conditioning}e). The compatible statement leaves the nominal fibril count close to the observation-only result, with the generating value of $19{,}950$ inside its $90\%$ interval. The conflicting statement shifts the median count to $18{,}700$ ($90\%$ interval $17{,}800$--$19{,}600$), excluding the generating value. The conditional posterior changes substantially even though the supplied separation is poorly supported by the observation-only posterior.

\subsection{What changes as humidity increases?}
\label{sec:app2-series}

\begin{figure}[t]
  \centering
  \includegraphics[width=\textwidth]{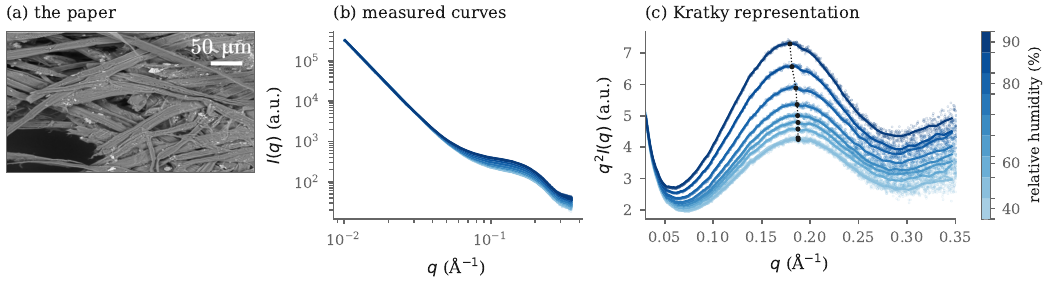}
  \caption{The measured series. (a)~Scanning electron micrograph of the
  paper, a network of cellulose fibres \citep[image by L.~Bengtsson;][]{bengtsson2023thesis};
  the SAXS window probes the cellulose fibrils inside the fibre walls, a few
  nanometres across. (b,c)~Nine SAXS curves of the same paper at relative
  humidities from 40\% (light) to 90\% (dark). Small dots are the measured
  points; lines are their averages in logarithmic $q$-bins, shown for
  readability. (b)~On logarithmic axes the curves nearly overlap. (c)~The same
  data in Kratky representation, $q^2I(q)$, a display transform that flattens
  the steep low-$q$ decay; black markers show the maximum of each curve.
  Inference uses the window $0.05$--$0.30$~\AA$^{-1}$.}
  \label{fig:app2-curves}
\end{figure}

We analyze nine SAXS measurements of the same paper specimen, taken between $40\%$ and $90\%$ RH (Fig.~\ref{fig:app2-curves}). Similar humidity series of paper materials have been analyzed by fitting analytical scattering models \citep{bjorn2025thesis}. The geometric mean scattering intensity over $0.05$--$0.30$~\AA$^{-1}$ increases by about $55\%$ between the driest and wettest states, with the largest relative changes around the Kratky peak near $q=0.18$~\AA$^{-1}$. The peak becomes more pronounced while its position remains between $0.178$ and $0.185$~\AA$^{-1}$ (Fig.~\ref{fig:app2-curves}c).

We restrict inference to $0.05$--$0.30$~\AA$^{-1}$ because the simulator cannot reproduce the steeper decay measured at lower $q$ without distorting the fibril-scale parameters (Appendix~\ref{app:instrument}). The simulated example shows what information is lost: excluding the low-$q$ region mainly broadens the posterior of the matrix correlation length, which describes larger-scale structure in the model (Section~\ref{sec:app2-sim}). A common intensity alignment is applied across the series. Posterior-predictive curves have a median root-mean-square discrepancy of about $0.02$ log$_{10}$ intensity units, exceeding the assumed measurement noise. A systematic residual remains at high $q$ and increases slightly for the wettest measurement.

\begin{figure}[t]
  \centering
  \includegraphics[width=0.95\textwidth]{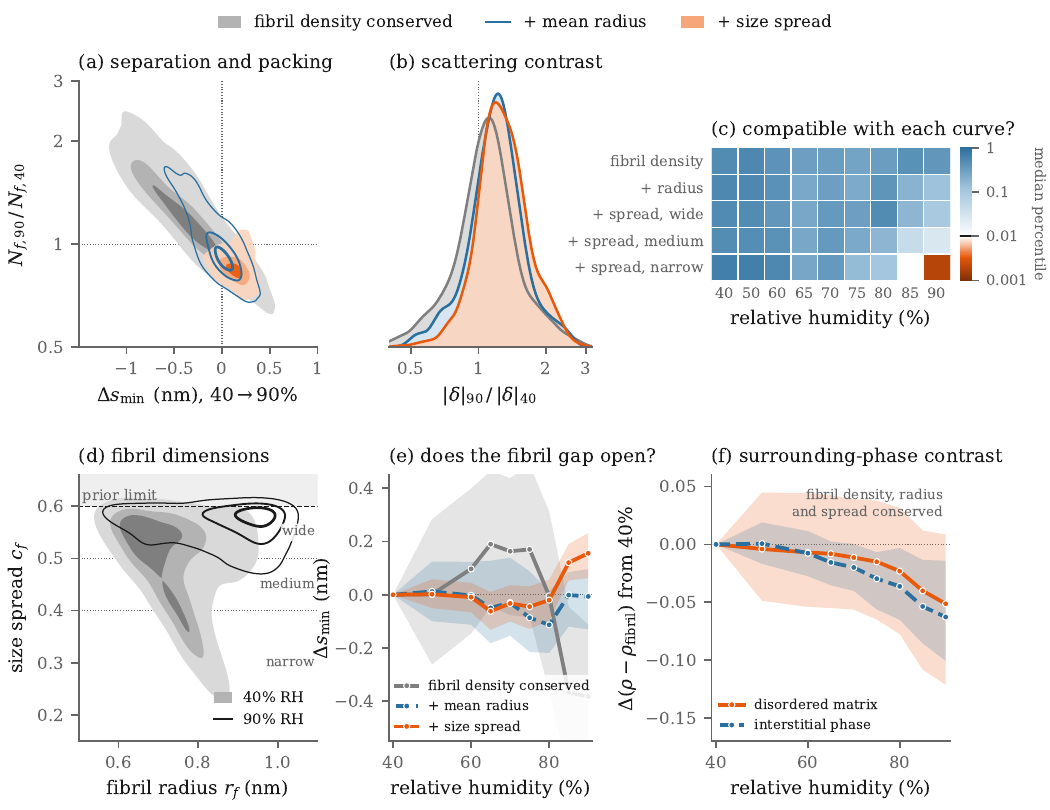}
  \caption{Inferred structural changes during hydration,
comparing observation-only inference with three assumptions
about conserved fibril properties: density (grey), density
and mean radius (blue), and density, radius and size spread
(orange). Density conservation gives results similar to
observation-only inference. Conserved values are drawn
jointly from the posterior at 40\% RH and supplied to
every measurement.
  \textbf{(a)}~Joint distribution of the change in minimum fibril separation and the ratio of nominal fibril counts between 40\% and 90\% RH; regions contain 20\%, 50\% and 90\% of the posterior mass.
  \textbf{(b)}~Ratio of scattering-contrast magnitudes between 90\% and 40\% RH.
  \textbf{(c)}~Median compatibility percentile of the supplied values against the observation-only posterior at each humidity (Section~\ref{sec:compat}). The density–radius–spread hypothesis is divided by supplied size spread: wide ($c_f\geq0.5$), medium ($0.4\leq c_f<0.5$), and narrow ($c_f<0.4$). The bar marks the flag threshold of 0.01.
  \textbf{(d)}~Observation-only joint posterior of fibril radius and size spread at 40\% (filled) and 90\% RH (contours); dotted lines mark the size-spread groups in (c), and shading marks the simulated upper limit.
  \textbf{(e)}~Change in minimum separation relative to 40\% RH under the three assumptions; markers indicate measured humidities and bands contain 50\% of the draws.
  \textbf{(f)}~Changes in matrix and interstitial phase densities relative to the fibrils, under the density–radius–spread assumption.}
  \label{fig:app2-humidity}
\end{figure}

The inferred changes in minimum fibril separation and nominal fibril count are broad and strongly anti-correlated ($r=-0.88$; Fig.~\ref{fig:app2-humidity}a), allowing both closer packing of more fibrils and wider separation of fewer. The nominal count $N_f$ acts as a packing-density parameter rather than a direct count of fibrils (Appendix~\ref{app:simulator}). The inferred fibril dimensions change little up to $80\%$ RH, but at $85$--$90\%$ the posterior shifts towards larger radii and greater size spread (Fig.~\ref{fig:app2-humidity}d). The size spread reaches $0.57$ ($90\%$ interval $0.47$--$0.60$), close to its simulated upper limit.

The observation-only inference suggests that the effective fibril size distribution broadens at high humidity. But do the measurements require this change, or could the scattering also be explained if fibril dimensions remain unchanged? We examine three increasingly restrictive assumptions: conservation of crystalline fibril density alone, density and mean radius, or density, radius and size spread. The first assumes that water does not enter crystalline cellulose. Conservation of fibril dimensions is less certain, since water uptake between fibrils \citep{penttila2021wateraccess} and moisture-induced fibril straining \citep{Salmen2021} have both been reported, and model fits to SAXS of a paper material indicate changes in both fibril diameter and interfibrillar distance with humidity \citep{bjorn2025thesis}. For each assumption, we draw $1{,}200$ joint samples from the posterior at $40\%$ RH. We hold the conserved parameters from each draw fixed across the humidity series, while inferring the remaining parameters separately from each measurement (Eq.~\eqref{eq:assumed-mixture}). The draws retain uncertainty about the values of the conserved parameters, but each draw assumes that those values remain unchanged across humidity. The later measurements do not update the assumed values.

Conserving fibril density alone has little effect on the inferred
geometry. With density and mean radius conserved, the median change
in minimum separation between $40\%$ and $90\%$ RH is $-0.01$~nm
($90\%$ interval $-0.45$ to $+0.26$~nm), while the size distribution
broadens (Fig.~\ref{fig:app2-humidity}a,~e, blue). Conserving size
spread as well shifts the median separation change to $+0.16$~nm
($-0.08$ to $+0.34$~nm) and reduces the nominal fibril count to
$85\%$ of its dry value (Fig.~\ref{fig:app2-humidity}a,~e, orange).
Both assumptions reproduce the measured curves similarly, despite
the different structural interpretations.

These conditional results show what follows from the conservation
assumptions, but not whether the supplied parameter values are
supported by the measurements. We therefore examine their
compatibility with the observation-only posterior
(Fig.~\ref{fig:app2-humidity}c). Few supplied values are flagged
when only fibril density is conserved. When the mean radius is
also conserved, $17\%$ of supplied draws are flagged at $90\%$
RH, compared with $1\%$ at $40\%$. Adding conservation of size
spread gives the lowest compatibility percentiles for narrow
distributions: at $90\%$ RH, the median is $0.002$ for spreads
below $0.4$, compared with $0.10$ for spreads above $0.5$.

The median scattering-contrast magnitude increases with humidity under all three assumptions, by a factor of $1.1$--$1.3$ at $90\%$ RH (Fig.~\ref{fig:app2-humidity}b). On the dominant contrast branch, where the interstitial phase is less dense than the fibrils ($88\%$ of draws), the median matrix and interstitial densities relative to the fibrils decrease from about $60\%$ RH, both by about $0.06$ at $90\%$ (Fig.~\ref{fig:app2-humidity}f). The posterior intervals are broad, and the alternative contrast branch shows no clear trend. The decreases are consistent with water uptake affecting the surrounding material, but do not establish how water is distributed between the phases.

The nearly unchanged scattering-peak position does not determine whether the minimum separation between fibrils changes during hydration. The inferred separation depends on the assumed fibril dimensions, while the broader effective size distribution at high humidity could reflect nonuniform swelling or other structural changes. The size-spread posterior also approaches its upper prior bound at high humidity, limiting its physical interpretation. Independent information about fibril dimensions would help distinguish the alternative explanations.

\section{Related work}
\label{sec:related}

\paragraph{Amortized simulation-based inference.}
Neural SBI learns reusable approximations to posteriors, likelihoods, or likelihood ratios from simulator draws \citep{cranmer2020frontier,papamakarios2016fast,greenberg2019apt,papamakarios2019snl,hermans2020nre,ardizzone2019inn,radev2020bayesflow}.
Conventional amortized posterior estimators typically infer a fixed set of parameters from each new observation.
Prior work on SAXS inversion of lipid nanoparticles has shown that near-identical scattering curves can correspond to distinct parameter modes, motivating posterior inference that represents this ambiguity rather than returning a single fitted solution \citep{bankestad2026saxs}.

\paragraph{Arbitrary conditioning.}
Arbitrary conditioning has been studied in VAEs \citep{ivanov2019vaeac}, normalizing flows \citep{li2020acflow}, and neural processes \citep{garnelo2018cnp,nguyen2022tnp,chang2025ace}.
Within SBI, Simformer uses masked diffusion to learn arbitrary conditional distributions of parameters and observations \citep{gloeckler2024allinone}, and OneFlowSBI answers such queries with masked flow matching, treating the observations as further coordinates of the joint \citep{nautiyal2026oneflowsbi}. Our setting differs in that the observation remains fixed while the supplied parameter information changes. We therefore encode the observation separately and use parameter tokens to represent each conditional query.

\paragraph{Uncertain conditioning.}
We use conditional flow matching
\citep{lipman2023flow,tong2023cfm,wildberger2023flowmatching} to learn conditional posteriors; score-based diffusion models have been used for SBI in the same role \citep{geffner2023compositional,sharrock2024sequential}.
Uncertain parameter statements are represented through a likelihood and marginalized rather than fixed, which corresponds to likelihood, or virtual, evidence in the sense of \citet{pearl1988probabilistic}.
Alternatively, samples from the observation-only posterior can be reweighted by the statement likelihood, but this becomes inefficient when few samples receive appreciable weight.
The model learns these conditional posteriors across parameter subsets and statement uncertainties. Assumed parameter distributions are treated differently from likelihood evidence: the former are propagated through exact conditionals without being updated by the current observation, whereas the latter update the distribution of the latent parameters. This distinction is related to Jeffrey's conditioning \citep{jeffrey1965logic}; \citet{chan2005revision} discuss its difference from likelihood evidence.

\paragraph{Checking supplied information.}
Our compatibility diagnostic is related in spirit to prior--data conflict checks \citep{evans2006conflict} and predictive model checks \citep{box1980sampling}.
Rather than checking whether the observed data are surprising under a model, we assess whether a supplied parameter statement is unusual relative to the observation-only posterior.

\paragraph{Fixed observation as context.}
Architectures such as prior-data fitted networks and the Perceiver separate a shared context from the quantities being predicted \citep{mueller2022pfn,hollmann2023tabpfn,jaegle2021perceiver}.
We use this principle for repeated inverse-problem queries, compressing the fixed observation into shared context while parameter tokens represent changing conditional questions.
We evaluate both conditional posterior accuracy and computational cost.

\paragraph{Evaluation without a reference posterior.}
Simulation-based calibration and coverage diagnostics assess whether posterior uncertainty has the expected frequency \citep{talts2018sbc,hermans2022crisis,lemos2023tarp}.
These checks do not by themselves determine whether the full joint conditional distribution is accurate.
We therefore combine coverage diagnostics with the conditional log score, a strictly proper scoring rule \citep{gneiting2007proper}, using held-out simulator draws rather than a reference posterior.

\section{Discussion}
\label{sec:discussion}

The trained model provides a reusable approximation to the inverse problem defined by the simulator and prior. Rather than solving the inverse problem separately for each observation or parameter constraint, it lets us examine different joint conditional distributions within the same learned model. This makes it possible to investigate parameter dependencies, alternative explanations, and how additional information could distinguish them across the simulator's parameter space.

An important outcome is understanding what the measurements cannot determine. Different regions of the joint posterior may correspond to distinct physical explanations, while parameter dependencies can reveal which properties are constrained only in combination. Supplying additional information can favor different explanations rather than simply reduce uncertainty. The cellulose analysis also shows why a conditional interpretation should not be mistaken for evidence supporting its assumptions. Compatibility assessment can identify conflicts between supplied information and the observation-only inference, but cannot establish the validity of a physical hypothesis.

These remaining ambiguities can help guide further experiments. Some are inherent to the forward model and cannot be resolved by measuring the same observable more precisely, while others arise from similar but distinguishable responses. Examining how the posterior changes with additional parameter information may help identify useful complementary measurements. A natural extension is experimental design, where candidate measurements or conditions are evaluated for their ability to distinguish competing explanations. This would require models of their possible outcomes, beyond the conditional queries considered here.

The inferred distributions remain limited by the simulator, prior, and posterior approximation. In the cellulose application, the simulator could not reproduce the measured scattering at low \(q\), so we excluded this region from inference. This may reflect missing structural variation at larger length scales, such as differences in fibril organization and packing. The model also simplifies the matrix and its interfaces. Including additional structural disorder or a more explicit description of components such as hemicellulose could improve agreement with the measurements. The large inferred size spread at high humidity may similarly compensate for variation not represented by the simulator. These possibilities remain to be tested. A more detailed simulator could explain more of the data, but would also introduce parameters whose identifiability would need to be assessed.

\section{Conclusion}
\label{sec:conclusion}
We developed an amortized conditional inference method for revisiting scientific measurements under changing parameter information and physical assumptions. The cellulose SAXS application shows how this can reveal alternative interpretations of the same measurements, while retaining uncertainty in supplied information and assessing conflicts with the observation-only inference. The method does not resolve ambiguities inherent in the measurement, but makes their dependence on additional information and assumptions explicit.

\section*{Data and code availability}

The code for model training and inference and the trained models are available at \url{https://github.com/mariabankestad/aisaxs-sbi}.
The experimental SAXS measurements are available from the authors upon
reasonable request.

\ack{We acknowledge MAX IV Laboratory for time on the ForMAX beamline under proposals 20221506 and 20240661, where the SAXS measurements were performed, and we thank Jonas Engqvist and Kim Nygård for coordinating the beamtime. Research conducted at MAX IV, a Swedish national user facility, is supported by the Swedish Research Council under contract 2018-07152, the Swedish Governmental Agency for Innovation Systems under contract 2018-04969, and Formas under contract 2019-02496. We acknowledge the EuroHPC Joint Undertaking for awarding this project access to the EuroHPC supercomputer LEONARDO, hosted by CINECA (Italy) and the LEONARDO consortium through an EuroHPC Development Access call (project EUHPC\_D34\_201).}

\funding{This work was funded by Vinnova through the project AI-SAXS: Decoding Structural Complexity with Intelligent Scattering Analysis (grant No.~2023-02701), with in-kind co-funding from the project partners.}

\clearpage

\appendix

\section{Conditional flow-matching details}
\label{app:conditional-flow}

\subsection{Masked conditional flow-matching objective}
\label{app:masked-objective}

For exact subset conditioning, let $\cmask\in\{0,1\}^d$ denote the mask, with
$c_i=1$ for an exactly supplied coordinate and $c_i=0$ for an unknown
coordinate. Given
\[
    \thetab_t=(1-t)\bm{\epsilon}+t\thetab,
    \qquad
    \bm{\epsilon}\sim\Ncal(\bm{0},\bm{I}),
\]
the state presented to the network is
\begin{equation}
    \widetilde{\thetab}
    =
    \cmask\odot\thetab
    +
    (\bm{1}-\cmask)\odot\thetab_t .
    \label{eq:masked-state-app}
\end{equation}
Exactly supplied coordinates remain fixed, while unknown coordinates follow the
flow path. The loss is evaluated only on the unknown coordinates:
\begin{equation}
\begin{aligned}
    \mathcal{L}(\phi)
    =
    \E_{t,\bm{\epsilon},(\thetab,\xb),\cmask}
    \Bigg[
        \frac{
        \big\|
        (\bm{1}-\cmask)\odot
        \big(
            \vb_\phi(\widetilde{\thetab},t,\xb,\cmask)
            -
            (\thetab-\bm{\epsilon})
        \big)
        \big\|^2
        }{\|\bm{1}-\cmask\|_1}
    \Bigg].
\end{aligned}
\label{eq:masked-loss}
\end{equation}
The normalization removes the dependence of the loss magnitude on the number
of unknown coordinates. During training, we draw the number of supplied
parameters uniformly from $0$ to $d-1$ and then choose a subset of that size
uniformly. The fully supplied case is excluded because there is nothing to
infer.

\begin{proposition}[Masked training recovers conditionals in the mask support]
\label{prop:masked}
Let $\cmask$ be drawn independently of $(\thetab,\xb)$ from a distribution
whose support contains the conditioning patterns of interest. For every mask
in that support and almost every $(\xb,\thknown)$ under the induced joint, the
population minimizer of \eqref{eq:masked-loss} over an unrestricted velocity
class is the conditional flow-matching field transporting a standard Gaussian
on the unknown coordinates to
\[
    p(\thunknown\mid\xb,\thknown).
\]
Thus a single population-optimal masked model recovers the joint conditional
posterior for every conditioning pattern represented during training.
\end{proposition}

This is a population result; finite-data accuracy can differ across
conditioning patterns. The proof is given in Appendix~\ref{app:proofs}.

\subsection{Conditioning on uncertain statements}
\label{app:uncertain-objective}

For uncertain supplied information, the reported value is generated as
\begin{equation}
    a_i=\theta_i+\sigma_i\eta_i,
    \qquad
    \eta_i\sim\Ncal(0,1),
    \label{eq:statement-corruption-app}
\end{equation}
so that $p(a_i\mid\theta_i,\sigma_i)$ defines an auxiliary likelihood for the
statement. The model receives $(a_i,\sigma_i)$ rather than the latent
$\theta_i$, which is marginalized as in
\eqref{eq:statement-marginal}.

\begin{proposition}[Training with uncertain statements recovers the target conditional]
\label{prop:soft}
Under the statement-generation model of
\eqref{eq:statement-corruption-app}, the population minimizer of the masked
flow-matching objective conditioned on the supplied statements recovers
\[
    p\!\left(
        \thetab_U
        \mid
        \xb,\bm{a}_T,\bm{\sigma}_T,
        \thetab_E=\bm{a}_E
    \right),
\]
with the uncertain latent values marginalized as in
\eqref{eq:statement-marginal}. Exact supplied values are handled by the
exact-conditioning mechanism.
\end{proposition}

Again, finite-data accuracy depends on the statement uncertainties and
conditioning patterns represented during training. The proof is given in
Appendix~\ref{app:proofs}.

\section{Proofs}
\label{app:proofs}

\begin{proof}[Proof of Proposition~\ref{prop:masked}]
The objective is an expectation over masks, and the network receives the mask,
observation, and supplied coordinates as inputs. Minimization therefore
decomposes pointwise over $(\cmask,\xb,\thknown)$. For a fixed conditioning
pattern, the objective on the unknown coordinates is the conditional
flow-matching regression for
$p(\thunknown\mid\xb,\thknown)$. Its $L^2$ minimizer is the marginal velocity
field generating that distribution \citep{lipman2023flow}.
\end{proof}

\begin{proof}[Proof of Proposition~\ref{prop:soft}]
The same argument applies with
$(\cmask,\xb,\bm{a},\bm{\sigma})$ as conditioning variables. For uncertain
coordinates, Bayes' rule reweights the observation-only posterior by the
statement likelihood,
\begin{equation}
    p(\thetab_T\mid\xb,\bm{a}_T,\bm{\sigma}_T,\thetab_E=\bm{a}_E)
    \propto
    p(\thetab_T\mid\xb,\thetab_E=\bm{a}_E)
    \prod_{i\in T} k(a_i\mid\theta_i,\sigma_i),
    \label{eq:statement-weight}
\end{equation}
where $k$ is the Gaussian statement likelihood. Exact coordinates in $E$ are
fixed at $\bm{a}_E$. The regression targets for $\thetab_U$ therefore follow
the corresponding mixture. Its velocity field is again the $L^2$ minimizer
\citep{lipman2023flow}, and marginalizing $\thetab_T$ gives
\eqref{eq:statement-marginal}.
\end{proof}

\section{Implementation details}
\label{app:impl}

We give the architecture, parameter transformations and training settings
used in the experiments, followed by the computational cost of repeated
inference. Unless stated otherwise, the settings in this section refer to the
controlled benchmarks of Section~\ref{sec:experiments}; the cellulose
application uses the configuration described in
Appendix~\ref{app:instrument}.

\paragraph{Architecture.}
The observation encoder is a three-layer MLP
$m\to128\to128\to K\cdot128$, reshaped into $K$ summary tokens with learned
positional embeddings. Parameter tokens encode parameter identity, value or
statement information, and role. Each of six transformer blocks applies
cross-attention to the observation summaries, self-attention among parameter
tokens, and a feed-forward layer. Flow time is injected through adaptive layer
normalization. All blocks use eight attention heads, and a linear head predicts
one velocity component per parameter token.

Since the observation does not change during the flow integration or between
repeated queries, its summary representation can be computed once and reused.
For the timing comparisons in Section~\ref{sec:experiments}, however, we
deliberately recompute the representation at every solver step, so the
reported speedups reflect the architecture itself rather than additional
savings from caching. The joint-attention baseline uses
matched-width and matched-depth self-attention over all $d+m$ tokens. The
no-attention baseline uses the same observation encoder but processes parameter
tokens independently.

\paragraph{Bounded parameters.}
Parameters in $[0,1]$ are mapped through an $\epsilon$-padded probit transform
and standardized per dimension using the training set. The flow operates in
this unconstrained space, and posterior samples are transformed back to the
original parameter space. For a uniform prior, the transformed prior is close
to the Gaussian base distribution.

\paragraph{Statement distribution and embedding.}
During training, each supplied parameter is exact with probability $0.7$.
Otherwise, its uncertainty is drawn log-uniformly from $[0.005,0.15]$ in
normalized units, and Gaussian noise with that standard deviation is added to
the true value. Statement values are not clipped to the parameter bounds.
Statements with uncertainties outside the training range are not accepted at
inference.
Exact and uncertain statements share the same value embedding, while
uncertainty is encoded separately. At zero uncertainty, the representation
reduces to exact conditioning. Statement values and uncertainties are defined
in normalized parameter units; the probit transform is used only for
network inputs. In a matched ablation, jointly embedding value and uncertainty
reduced the zero-uncertainty conditional log score by $0.58$ nat. Sharing the
value embedding recovered $0.25$ nat, and the statement-training distribution
a further $0.32$ nat. Table~\ref{tab:gauss-metrics} compares the two designs
at nonzero uncertainty on the Gaussian task; C2ST does not clearly separate
them, whereas their conditional log scores differ by $0.42 \pm 0.08$ nat.

\begin{table}[t]
\centering
\caption{C2ST and conditional log score for two statement embeddings on the Gaussian
task. C2ST is evaluated at $\sigma=0.05$, where $0.5$ indicates
chance level against the analytic posterior. The conditional log score averages the model log-density at the
held-out parameters over $500$ conditioning patterns; higher is better.}
\label{tab:gauss-metrics}
\small
\setlength{\tabcolsep}{5pt}
\begin{tabular}{lcc}
\toprule
statement embedding & C2ST & conditional log score (nat) \\
\midrule
shared-value & $0.523 \pm 0.023$ & $2.90 \pm 0.06$ \\
joint value--uncertainty & $0.509 \pm 0.022$ & $2.48 \pm 0.05$ \\
\midrule
difference & $0.014$ & $0.42 \pm 0.08$ \\
\bottomrule
\end{tabular}
\end{table}

\paragraph{Training.}
The models in Section~\ref{sec:experiments} are trained with AdamW, weight
decay on weight matrices, gradient clipping, and a cosine learning-rate
schedule for at most $250$ epochs. Early
stopping uses validation loss with patience $60$. An exponential moving average
of the weights with decay $0.999$ is used for validation, model selection, and
all reported results. All compared architectures use the same training budget
and settings. Without the moving average, $90\%$ coverage on the SAXS task
falls from about $0.90$ to $0.82$, with a several-fold increase in marginal
calibration error.

\paragraph{Computational cost.}
If the observation encoding is cached, $m$ enters only through the encoder,
while each of the $S$ solver steps operates on $d$ parameter tokens and $K$
summary tokens:
\begin{equation}
  \text{compressed: } \mathcal{O}(m) + S\,\mathcal{O}(d^2+dK),
  \qquad
  \text{joint: } S\,\mathcal{O}((d+m)^2).
  \label{eq:cost}
\end{equation}
These are per-layer scaling comparisons rather than runtime predictions.
The reported timings exclude the additional benefit of caching, as noted
above.

\section{Evaluation protocols}
\label{app:evaluation}

\paragraph{Shared evaluation.}
Conditional calibration and coverage are evaluated on $1{,}000$ test
conditionals with $1{,}000$ posterior samples and $50$ solver steps each. We
sample the number of supplied parameters uniformly and then choose a subset
uniformly among subsets of that size. Curves are drawn with replacement from
the held-out set under a fixed seed, so all models use the same curves,
conditioning patterns, solver settings, and parameter transforms.

The conditional log density is computed by integrating the exact Jacobian trace
of the velocity field backward through the flow using $30$ deterministic
midpoint steps. For statement-conditioned models, supplied coordinates are
evaluated as zero-uncertainty statements. Densities are evaluated in transformed
coordinates; the fixed transform Jacobian is common to all models and therefore
cancels in comparisons. Contradictory statements (Section~\ref{sec:experiments})
are evaluated separately.

\paragraph{Metric definitions.}
Marginal calibration error is the mean absolute deviation of the empirical rank
distribution from uniformity, averaged over parameters and nominal levels.
$\mathrm{cov}_{90}$ is the fraction of true values inside central $90\%$
marginal credible intervals, averaged over parameters and conditionals. TARP
error is the mean absolute deviation between expected joint coverage and nominal
coverage across levels, using isotropic reference points centered at the
posterior mean.

\needspace{8\baselineskip}
\paragraph{Conditional log score.}
\label{app:conditional-score}

\begin{proposition}[The conditional log score is strictly proper]
\label{prop:proper}
For model conditionals $q$ and true conditionals $p$, the score in
\eqref{eq:logprob-score} satisfies
\begin{equation}
\mathcal{S}(q)
=
\mathcal{S}(p)
-
\E_{\xb,\cmask,\thknown^*}
\left[
\mathrm{KL}\!\left(
p(\cdot\mid\xb,\thknown^*)
\,\middle\|\,
q(\cdot\mid\xb,\thknown^*)
\right)
\right].
\label{eq:proper-score}
\end{equation}
Thus $\mathcal{S}$ is strictly proper and is uniquely maximized by the true
conditional distribution.
\end{proposition}

\begin{proof}
Conditional on $(\xb,\cmask,\thknown^*)$, the true unknown coordinates follow
$p(\cdot\mid\xb,\thknown^*)$, so
\[
    \E_p[\log q]
    =
    -\mathrm{H}(p)-\mathrm{KL}(p\|q).
\]
Taking the outer expectation gives \eqref{eq:proper-score}. Uniqueness follows
because $\mathrm{KL}(p\|q)=0$ if and only if $q=p$ almost everywhere.
\end{proof}

\paragraph{C2ST protocol.}
For the Gaussian validation, each C2ST compares $2{,}000$ samples from the
learned conditional with $2{,}000$ samples from the corresponding analytic
conditional. We evaluate $12$ choices of supplied parameters and values at each
uncertainty level. Each test uses a two-layer MLP classifier with five-fold
cross-validation. The classifier is retrained for every conditioning pattern,
and the reported variation is the standard deviation across patterns.

\paragraph{Timing protocol.}
A query includes observation encoding, ODE integration, and on-device sampling
of $1{,}000$ draws using $50$ fixed Euler steps in float32 with eager PyTorch.
Fixed DCT preprocessing and host transfers are excluded. Timings are medians of
three repeats after one warm-up pass with device synchronization. GPU timings
use one A100-SXM-64GB; CPU timings use an Ice Lake Xeon node restricted to
eight threads. The observation-dimension benchmark varies the number of DCT
coefficients while keeping the encoder architecture fixed. Random inputs are
used because forward-pass cost does not depend on the model weights.

\paragraph{Compatibility percentile.}
For a statement $(\bm{a},\bm{\sigma})$ on coordinates $k$, we draw
$N$ samples
$\thknown^{(n)}\sim q(\thknown\mid\xb)$ and compute
\begin{equation}
s(\bm{a})
=
\log\frac{1}{N}
\sum_n
\prod_{i\in k}
\Ncal\!\left(
a_i;\theta_i^{(n)},\tilde{\sigma}_i^2
\right),
\end{equation}
where $\tilde{\sigma}_i=\max(\sigma_i,0.01)$, so exact statements are evaluated
with a small nonzero width. Reference statements are generated from the same
posterior,
\(
a_i^{(r)}
=
\theta_i^{(r)}
+
\tilde{\sigma}_i\epsilon,
\)
using $R$ posterior draws. The compatibility percentile is the fraction of
reference scores no greater than $s(\bm{a})$. The reported analyses use
$N\ge1{,}000$ and $R\ge256$. When the supplied statement is generated by the
same procedure as the reference statements, its rank is approximately uniform,
apart from finite-sample effects. One such effect matters for statements on
several coordinates: reference statements jittered from the same $N$ samples
that score them lie close to one of their own scoring samples and score
higher than an independent posterior draw, which flagged about $10\%$ of
posterior draws for the three-coordinate statements of
Section~\ref{sec:app2-series}. The humidity-series percentiles therefore
score with one half of the posterior samples and generate the reference
statements from the other half; at the reference humidity $0$--$1.2\%$ of
posterior draws are then flagged (Appendix~\ref{app:instrument}).

\paragraph{Solver convergence.}
For the Gaussian validation, statement-conditioning C2ST values are insensitive
to the tested solver settings. At the largest uncertainty, midpoint integration
with $10$, $25$, and $50$ steps and Euler integration with $50$ steps agree
within $0.005$, well below the variation across conditioning patterns
($\pm0.05$). Reported values use $50$-step midpoint integration.

\section{The cellulose scattering simulator}
\label{app:simulator}

Section~\ref{sec:exp-setup} gives the physical relation to established
cellulose SAXS descriptions. Here we specify the real-space construction and
parameterization used in the simulations. The simulated structure is a
cross-section perpendicular to the fibril direction. It can equivalently be
regarded as a three-dimensional structure obtained by extending this
cross-section along the third dimension, corresponding to parallel straight
fibrils. The scattering in the plane perpendicular to the fibril direction
therefore has the same \(q\)-dependence as that calculated from the
two-dimensional cross-section, apart from an overall scale factor
\citep{stribeck2007slicetheorem}, analogous to the infinitely long-cylinder
approximation used in WoodSAS.

\paragraph{Forward model.}
A Gaussian random field with Mat\'ern-type spectral density, inverse
correlation-length parameter $\kappa$, and fixed smoothness is thresholded so
that the disordered-matrix phase occupies the fraction $\phi$ of the
cross-section. Hexagonal fibril cross-sections are then placed by random
sequential addition outside the matrix, with a minimum gap $s_{\min}$ between
neighbouring fibrils, so their centers are not constrained to a lattice. The
remaining area forms the interstitial phase, and the three phases partition
the cross-section (Fig.~\ref{fig:simulator}). Following
\citep{schmidt2007simulation,roding2022,bankestad2026saxs}, the scattering
intensity is the squared magnitude of the Fourier transform of the
scattering-length-density-weighted phase map. We average over six independent
realizations of the random field and fibril placement, apply light Gaussian
smoothing, and azimuthally average onto $1{,}000$ $q$ points, which are mapped
to \AA$^{-1}$ with the fixed $q$-calibration of the collaborating group. This
six-realization average defines the forward map.

\paragraph{Two configurations.}
The comparisons in Section~\ref{sec:experiments} use a simplified
configuration on a $1024\times1024$ grid in which all fibrils have the same
size, set by an inverse scale $a_f$. Candidate fibrils are proposed across the
whole cross-section and rejected if they overlap the matrix, and their outlines
are rasterized pixel by pixel, so their apparent areas vary slightly with
sub-pixel placement.

The application in Section~\ref{sec:app2} uses a more
detailed configuration on a $2048\times2048$ grid covering a square of side
$743.5$~nm. Fibril radii (hexagon circumradii) follow a lognormal
distribution with mean $r_f$ and coefficient of variation $c_f$, truncated to
$0.5$--$5$~nm. Fibrils are
proposed only where they fit outside the matrix, with the requested number
scaled by the available area, and each fibril is rasterized by its fractional
pixel coverage, so the curve changes continuously with fibril size. At this
resolution the curves agree with a $4096\times4096$ grid to within
$0.025$ decades below $0.22$~\AA$^{-1}$; this check does not cover the
highest $q$ of the measurement window. Both configurations share the
observation representation, the noise model, and the training objective.

\paragraph{Parameters and notation.}
Table~\ref{tab:params} lists the sampled parameters of both configurations.
Two quantities used in the application are derived from them: the nominal
fibril count $N_f=n_r a_f$, where $a_f=3.7175~\mathrm{nm}/r_f$ in the
application configuration, and the matrix correlation length
$\ell_m=1/\kappa$, in simulation units. Phase densities are relative
scattering-length densities.

\paragraph{Nominal fibril count.}
$N_f$ is the number of fibrils requested for one simulated cross-section
(the reference area) as if no matrix were present, not the number realized. In
the application configuration, the number of candidates proposed is $N_f$
times the fraction of the area where a fibril of mean size fits outside the
matrix, and candidates that violate the minimum separation are rejected, so the
realized count is smaller than $N_f$. We therefore report and interpret $N_f$ as a nominal model
parameter that acts as a packing density, not as a direct count of fibrils in
the material.

\begin{table}[t]
  \caption{Sampled forward-model parameters and prior ranges for the two
  simulator configurations. Ranges marked log are sampled uniformly in the
  logarithm. Derived quantities: nominal fibril count $N_f=n_r a_f$, matrix
  correlation length $\ell_m=1/\kappa$, and contrast magnitude
  $\lvert\bm{\delta}\rvert$.}
  \label{tab:params}
  \footnotesize
  \setlength{\tabcolsep}{3pt}
  \begin{tabular}{lllll}
    \toprule
    symbol & quantity & simplified (Sec.~\ref{sec:experiments}) & application (Sec.~\ref{sec:app2}) & code name \\
    \midrule
    $a_f$, $r_f$ & fibril size & $a_f\in[2,4]$ & $r_f\in[0.5,2.5]$~nm, log & \texttt{tube\_scaling}, \texttt{radius\_nm} \\
    $c_f$ & fibril size spread & -- & $[0,0.6]$ & \texttt{size\_cv} \\
    $n_r$ & fibril count factor & $[7500,10000]$ & $[1000,15000]$, log & \texttt{no\_rods\_per\_unit} \\
    $s_{\min}$ & min.\ fibril separation & $[0,1]$, sim.\ units & $[0,3]$~nm & \texttt{min\_space} \\
    $\phi$ & disordered-matrix volume fraction & $[0.2,0.5]$ & $[0.05,0.9]$ & \texttt{volfrac\_grf} \\
    $\kappa$ & inverse matrix corr.\ length & $[0.75,2.5]$ & $[0.25,3]$, log & \texttt{kappa\_sc} \\
    $\rho_{cry}$ & fibril phase density & $[0.7,1.0]$ & $[0.7,1.0]$ & \texttt{SLD\_tubes} \\
    $\rho_{m}$ & matrix phase density & $[0.5,0.99]$ & $[0.5,0.99]$ & \texttt{SLD\_grf\_1} \\
    $\rho_{int}$ & interstitial phase density & $[0.5,0.99]$ & $[0.5,0.99]$ & \texttt{SLD\_grf\_2} \\
    \bottomrule
  \end{tabular}
\end{table}

\paragraph{Contrast form.}
With $\delta_1=\rho_m-\rho_{cry}$ and $\delta_2=\rho_{int}-\rho_{cry}$,
Eq.~\eqref{eq:contrast-form} written out reads
\begin{equation}
I(q)
=
\delta_1^2S_{11}(q)
+
\delta_2^2S_{22}(q)
+
2\delta_1\delta_2S_{12}(q),
\label{eq:contrast-components}
\end{equation}
where $S_{ij}$ depend only on geometry and are averaged over the six
realizations. Thus three phase densities enter the simulator, but only two
density contrasts affect the scattering. The common offset
\[
(\rho_{cry}+\rho_{m}+\rho_{int})/\sqrt{3}
\]
is therefore structurally non-identifiable, and so is the joint sign of the
two contrasts (Fig.~\ref{fig:app2-invariance}a).

\paragraph{Data generation.}
For a fixed geometry, the three radially averaged cross-spectra
$S_{11}$, $S_{22}$, and $S_{12}$ determine the curve for any phase-density
triple. We verified this representation against direct simulation to relative
error below $10^{-4}$ across the prior, including near-zero contrast. Phase
densities are therefore redrawn for each batch and curves synthesized on the
fly from stored cross-spectra. The simplified models are trained on $65{,}536$
geometries from one scrambled Sobol sequence; validation and test sets use
disjoint Sobol streams with $512$ and $1{,}024$ geometries and fixed density
draws, so all models in Section~\ref{sec:experiments} are evaluated on the same
curves. The application models are trained on $131{,}072$ geometries and
tested on $1{,}024$ held-out geometries.

\section{Cellulose inference model and calibration}
\label{app:instrument}

\paragraph{Training.}
The application models use the compressed architecture with
$d_{\mathrm{model}}=192$, $K=8$, and six layers, and are trained on the
application configuration (Appendix~\ref{app:simulator}) with independently
sampled base noise. Each batch mixes nine $q$-windows,
$q_{lo}\in\{0.01,0.02,\ldots,0.08,0.10\}$~\AA$^{-1}$ with
$q_{hi}=0.30$~\AA$^{-1}$, so that one model serves both the full simulated
range and the measurement window. Measurement noise follows the $q$-dependence
of the measured curves, with an amplitude parameter drawn from
$[0.0007,0.010]$; the measured curves correspond to $0.0029$, the level used
for all inference in Section~\ref{sec:app2}. Three replicas are trained with
different seeds for $1{,}400$ epochs with a cosine learning-rate schedule and an
exponential moving average of the weights; the final training and validation
losses differ by about $1.5\%$.

\paragraph{Calibration on held-out simulations.}
On $1{,}024$ held-out test geometries with noise at the level of the measured
curves, the observation-only posteriors of each replica have empirical
coverage $0.90$ for central $90\%$ intervals and $0.50$ for central $50\%$
intervals, averaged over parameters. Pooling the three replicas gives $0.913$
and $0.522$, and the root-mean-square error of the pooled posterior mean is
$0.186$ in normalized units. These are averages over parameters and test
observations; they do not guarantee equal calibration for each parameter or
for unusual observations. We also check whether averaging the inferred
posterior over held-out observations approximately recovers the prior
distribution, as expected for a correctly specified inference procedure. The
mean mass per tenth of the range is
$0.096$--$0.106$ (prior $0.100$), and the outer $2.5\%$ at each bound holds
$0.023$--$0.027$ (prior $0.025$); with three parameters supplied exactly, the
largest deviation per tenth is $0.020$ (standard error about $0.012$). This
checks the aggregate distribution of posterior mass, not coverage for
individual observations near the prior bounds.

Results with $25$, $50$, and $100$ midpoint solver steps agree. No
calibration correction is applied.

\paragraph{Measured curves.}
The measurements are scanning SAXS \citep{bunk2009} of one paper specimen,
a network of cellulose fibres \citep{bengtsson2023thesis}, recorded at the ForMAX beamline, MAX IV Laboratory
($16$~keV; EIGER2 X 4M detector at $2.96$~m; $100\times100$~$\mu$m$^2$
beam; $50\times60$ grid with $100$~$\mu$m steps). The specimen was
conditioned for at least one hour at each relative humidity ($40$, $50$,
$60$, $65$, $70$, $75$, $80$, $85$, and $90\%$; ProUmid MHG-100 humidity
generator). The curve at each humidity is the sum over the same $2{,}196$
on-sample positions of the scan.

\paragraph{Measurement window.}
The measured curves are analyzed on $0.05$--$0.30$~\AA$^{-1}$. Between
$0.02$ and $0.05$~\AA$^{-1}$ they decay with a slope of $-3.6$ on
logarithmic axes at $40\%$ humidity and $-3.4$ at $90\%$, whereas posterior
draws re-simulated on the full simulator grid give $-2.8$ to $-3.0$ at every
humidity (root-mean-square misfit $0.11$--$0.14$ decades there, against
$0.013$--$0.024$ inside the window). Conditioning
on $0.03$--$0.30$ or $0.02$--$0.30$~\AA$^{-1}$ instead halves the misfit
below the window but does so by moving the fibril-scale parameters: at
$40\%$ the posterior median radius goes from $0.71$ to $0.90$--$0.95$~nm,
the size spread from $0.54$ to $0.08$--$0.18$, the minimum separation from
$0.7$ to $1.2$~nm and the nominal count from $7{,}400$ to
$3{,}500$--$4{,}400$, while the misfit inside the window grows from $0.018$
to $0.03$ decades. The oscillatory matrix spectrum of the full simulator,
absent from the trained configuration, steepens the decay only when the
matrix correlation length is also changed, which raises the window misfit to
$0.06$ decades or more. The low-$q$ part of the simulated example
(Section~\ref{sec:app2-sim}) is used only to show what such data would add.

\paragraph{Intensity alignment.}
The measured intensities are on a relative scale, and the nine curves share
one unknown scale. Before inference, all nine are therefore shifted by the
same amount $\Delta=-1.207$ in $\log_{10}$ intensity, which places the series
at the centre of the training distribution of intensity levels on the
measurement window. The alignment is a working choice, not a physical
calibration. It changes only the overall intensity level seen by the model,
not the curve shapes or the intensity ratios between humidities. Because the
intensity level carries information about the contrast magnitude, the absolute
values of the inferred contrasts, and through posterior dependences other
parameters, depend on this choice; Section~\ref{sec:app2} therefore reports
changes between humidities under one common alignment.\looseness-1

\paragraph{Agreement with measured curves.}
For each measured curve, $48$ posterior draws were simulated without noise
(six-realization average) and compared with the measurement after the same
intensity alignment. The median root-mean-square difference is
$0.017$--$0.026$ decades across the nine humidities, and $0.013$--$0.018$
decades after the mean of each $\log_{10}$ curve is removed. The simulated
curves lie above the measurements at the highest $q$, by up to $0.03$ decades
between $0.24$ and $0.30$~\AA$^{-1}$. Under the hypotheses of
Section~\ref{sec:app2-series} the differences are similar ($0.018$--$0.024$
decades with the radius and size spread conserved; $0.020$ and $0.023$
decades at $40$ and $90\%$ with the density conserved as well). The
three-phase simulator is therefore an approximate description of the measured
curves, and the inferred parameters are interpreted within it.

\paragraph{Pooled reporting.}
Section~\ref{sec:app2} pools equal numbers of posterior samples from the three
replicas, to reduce the dependence on an individual training run; the pooled
samples are not a calibrated representation of model-approximation error.
Replica-to-replica variation is reported where it matters, as for the
compatibility percentiles, which are computed per replica.

\paragraph{Hypotheses about the fibrils.}
Section~\ref{sec:app2-series} supplies to all nine curves the fibril density
$\rho_{cry}$; the density and the mean radius $r_f$; or the density, radius
and size spread $c_f$ (Fig.~\ref{fig:app2-humidity}). The supplied values are
$1{,}200$ joint draws from the pooled observation-only posterior of the
$40\%$ curve, equal numbers from each replica, the same draws for every
hypothesis restricted to the coordinates it fixes. The draws are assigned to
the three replicas in turn, and each is supplied exactly, by its replica, to
all nine curves including the $40\%$ curve (Eq.~\eqref{eq:assumed-mixture});
the remaining parameters are sampled once per draw and curve from the
conditional posterior with $25$ midpoint steps ($50$ steps move the medians
of the reported changes by at most $0.01$~nm and $2\%$). Changes relative to
$40\%$ are computed within each draw, so the dependence between humidities
comes from the shared draw and not from joint inference over the series. In
the measurement-only analysis each replica samples every curve independently
and changes are formed within replica. Highest-density regions in
Fig.~\ref{fig:app2-humidity} are Gaussian kernel density estimates of these
samples. Table~\ref{tab:series-hypotheses} lists the changes between $40$
and $90\%$ under every combination examined, including three not shown in
Section~\ref{sec:app2-series}: the radius alone, the size spread alone, and
the fibril dimensions together with the minimum separation.

\begin{table}[t]
\centering
\small
\caption{Changes between $40$ and $90\%$ relative humidity under each
hypothesis: median and $90\%$ interval over $1{,}200$ supplied draws
(measurement only: $1{,}200$ independent samples). $\Delta s_{\min}$ is the
change in minimum fibril separation, $N_f$ the nominal fibril count,
$|\delta|$ the scattering-contrast magnitude and $\Delta c_f$ the change in
size spread; a dash marks a conserved quantity. The second, fourth and
seventh rows are the three hypotheses of Fig.~\ref{fig:app2-humidity}.}
\label{tab:series-hypotheses}
\setlength{\tabcolsep}{5pt}
\begin{tabular}{lcccc}
\toprule
supplied to all curves & $\Delta s_{\min}$ (nm) & $N_{f,90}/N_{f,40}$ & $|\delta|_{90}/|\delta|_{40}$ & $\Delta c_f$ \\
\midrule
none & $-0.39$ ($-0.98$, $0.41$) & $1.23$ ($0.74$, $2.16$) & $1.13$ ($0.56$, $2.21$) & $0.06$ ($-0.05$, $0.28$) \\
$\rho_{cry}$ & $-0.38$ ($-0.96$, $0.38$) & $1.23$ ($0.74$, $2.06$) & $1.09$ ($0.60$, $1.90$) & $0.05$ ($-0.05$, $0.26$) \\
$r_f$ & $0.00$ ($-0.48$, $0.26$) & $0.94$ ($0.78$, $1.59$) & $1.31$ ($0.68$, $2.31$) & $0.04$ ($-0.04$, $0.24$) \\
$\rho_{cry}$, $r_f$ & $-0.01$ ($-0.45$, $0.26$) & $0.93$ ($0.78$, $1.57$) & $1.20$ ($0.70$, $1.97$) & $0.04$ ($-0.04$, $0.24$) \\
$c_f$ & $-0.18$ ($-0.95$, $0.57$) & $1.02$ ($0.67$, $1.64$) & $1.23$ ($0.63$, $2.36$) & -- \\
$r_f$, $c_f$ & $0.16$ ($-0.06$, $0.36$) & $0.86$ ($0.73$, $1.06$) & $1.49$ ($0.88$, $2.38$) & -- \\
$\rho_{cry}$, $r_f$, $c_f$ & $0.16$ ($-0.08$, $0.34$) & $0.85$ ($0.74$, $1.04$) & $1.26$ ($0.84$, $2.04$) & -- \\
$r_f$, $c_f$, $s_{\min}$ & -- & $0.88$ ($0.78$, $1.01$) & $1.46$ ($0.90$, $2.42$) & -- \\
\bottomrule
\end{tabular}
\end{table}

Conserving the mean radius removes most of the decrease in separation that
the measurement-only posterior admits, because that posterior is in part a
trade-off between separation and radius: larger fibrils packed more closely
and fibrils of the dry size spaced as before both explain the wet
measurements, and the supplied radius excludes the first. Conserving the size
spread alone does not have this effect.

Conserving the fibril density leaves the geometry unchanged, because the
scattering depends only on density differences
(Eq.~\eqref{eq:contrast-components}); it fixes the offset of the three phase
densities and acts only through the prior. With $\rho_{cry}$ held near its
$40\%$ value the other two densities must stay within the simulated range,
which limits how far the contrast magnitude can grow: $1.26$ instead of
$1.49$ at $90\%$ with the radius and spread conserved. The relative
densities of Fig.~\ref{fig:app2-humidity}f are reported on the branch in
which the interstitial phase is less dense than the fibrils at $40\%$, which
holds $88\%$ of the draws; the matrix is also less dense than the fibrils in
only $57\%$. On the reflected branch neither relative density shows a clear
change. On the dominant branch the matrix and interstitial densities
relative to the fibrils fall by $0.06$ ($90\%$ interval $-0.39$ to $0.24$;
posterior probability of a decrease $0.77$) and $0.06$ ($-0.14$ to $0.39$;
$0.84$) at $90\%$. The median of their difference lies within $0.01$ of zero
at every humidity, but individual draws spread over about $\pm0.4$, so the
two changes are similar in the median only and neither is tightly
constrained. Without the density supplied the same quantities fall by $0.10$
and $0.08$ (probabilities $0.84$ and $0.91$); the difference is the prior
effect above.

\paragraph{Compatibility of the supplied values.}
Fig.~\ref{fig:app2-humidity}c scores each supplied draw against the
observation-only posterior of each curve with the compatibility percentile of
Section~\ref{sec:compat}, restricted to the coordinates the hypothesis fixes
($N=R=4{,}000$ per replica, $50$ midpoint steps, width $0.01$ in normalized
units), using the split construction of Appendix~\ref{app:evaluation}. A
supplied draw is a joint posterior draw only for the replica that produced
it, so each draw is scored by that replica, and the levels that include the
density use the $406$ draws whose dimensions and density come from the same
replica. At the reference humidity $0$--$1.2\%$ of the draws are flagged at
every level, as expected for draws from the posterior itself, and up to
$80\%$ humidity no level flags more than $10\%$. At $90\%$ the median
percentile and the flagged fraction are $0.37$ and $1\%$ with the density,
$0.13$ and $17\%$ with the radius added, and $0.03$ and $39\%$ with the
spread added. For the last, the conflict depends on the supplied spread:
$0.10$ and $23\%$ for spreads of $0.5$ and above ($197$ draws), $0.03$ and
$39\%$ between $0.4$ and $0.5$ ($122$), and $0.002$ and $74\%$ below $0.4$
($87$). Conserving the spread with the radius free gives $0.09$ and $26\%$.
At $85\%$ the pattern is the same with smaller fractions ($1$, $9$ and
$27\%$ for the three levels). The strongest conflict therefore comes from
narrow conserved size distributions, not from the conserved mean radius.

\paragraph{Compatibility example.}
The compatibility percentiles in Fig.~\ref{fig:app2-conditioning}d use
$N=R=4{,}000$ per replica and the definition in
Appendix~\ref{app:evaluation}; the conditional posteriors in
Fig.~\ref{fig:app2-conditioning}e use $1{,}000$ samples per replica and $50$
midpoint steps.

\end{document}